\documentclass[sigplan,10pt]{acmart}
\input{latexdiffp}
\renewcommand\footnotetextcopyrightpermission[1]{} 
\usepackage{natbib,hyperref}
\usepackage[normalem]{ulem} 

\usepackage{balance} 
\usepackage{array, boldline, rotating} 
\usepackage{color, colortbl} 
\definecolor{tabcol}{rgb}{0.7,0.8,1}
\usepackage{multirow}
\newcommand\DSL{sparse fusion}
\newcommand\SF{Multi-Sparse DAG Partitioning}
\newcommand\algo{MSP}
\newcommand\sfu{multi-sparse DAG partitioning}

\newcommand{\lastm}[1]{\iffalse\sout{#1}\fi}
\newcommand{\hkaz}[1]{\iffalse\sout{#1}\fi}

\newcommand{\kazed}[1]{\textcolor{black}{#1}}
\newcommand{\zedka}[1]{\textcolor{black}{#1}}
\newcommand{\cutc}[1]{\iffalse\textcolor{yellow}{#1}\fi}

\acmYear{2022}
\acmISBN{} 
\acmDOI{} 
\startPage{1}

\setcopyright{none}

\usepackage{booktabs}   
\usepackage{caption} 
\usepackage{subfig} 

\usepackage{listings}
\definecolor{commentgreen}{RGB}{2,112,10}
\definecolor{grayLine}{RGB}{100,100,100}
\definecolor{annotation}{rgb}{0.45, 0.31, 0.59}
\definecolor{darklavender}{rgb}{0.45, 0.31, 0.59}
\definecolor{asparagus}{rgb}{0.12, 0.3, 0.17}
\definecolor{brightmaroon}{rgb}{0.76, 0.13, 0.28}
\definecolor{bondiblue}{rgb}{0.0, 0.58, 0.71}
\newcommand{\kazem}[1]{\iffalse\textbf{\textcolor{blue}{[Kazem: #1]}}\fi}
\newcommand{\maryam}[1]{\iffalse\textbf{\textcolor{orange}{[Maryam: #1]}}\fi}

\newcommand{\rebt}[1]{{#1}}

\newcommand{\rebtout}[1]{\iffalse{\sout{#1}}\fi}

\usepackage{amsmath} 

\usepackage{algorithmic}
\usepackage[linesnumbered,lined,boxed,ruled,vlined]{algorithm2e}

\newsavebox{\mylistingbox}

\definecolor{commentgreen}{RGB}{2,112,10}

\SetCommentSty{mycommfont}

\begin{document}

\title{ Composing Loop-carried Dependence with Other Loops
}        



\author{Kazem Cheshmi}
\orcid{nnnn-nnnn-nnnn-nnnn}             
\affiliation{
  \institution{University of Toronto}            
  \city{Toronto}
  \country{Canada}                    
}
\email{kazem@cs.toronto.edu}          

\author{Michelle Mills Strout}
\affiliation{
  \institution{University of Arizona}           
  \city{Tucson}
  \country{USA}                   
}
\email{mstrout@cs.arizona.edu}         

\author{Maryam Mehri Dehnavi}
\affiliation{
  \institution{University of Toronto}           
  \city{Toronto}
  \country{Canada}                   
}
\email{mmehride@cs.toronto.edu}         

\begin{abstract}

Sparse fusion is a compile-time loop transformation and runtime scheduling implemented as a domain-specific code generator. Sparse fusion generates efficient parallel code for the combination of two sparse matrix kernels where at least one of the kernels has loop-carried dependencies. 
Available implementations optimize individual sparse kernels. 
When optimized separately, the irregular dependence patterns of sparse kernels create synchronization overheads  and load imbalance, and their irregular memory access patterns result in inefficient cache usage, which reduces parallel efficiency.  
Sparse fusion uses a novel inspection strategy  with code transformations to generate parallel fused code for sparse kernel combinations that is optimized for  data locality and load balance.  Code generated by Sparse fusion outperforms the existing implementations  ParSy and  MKL on average 1.6$\times$ and 5.1$\times$ respectively and outperforms the LBC and DAGP coarsening strategies applied to a fused data dependence graph on average 5.1$\times$ and 7.2$\times$ respectively for various kernel combinations. 
\end{abstract}



\maketitle
\pagestyle{plain}
\section{Introduction}

\begin{figure}
   \includegraphics[width=0.41\textwidth]{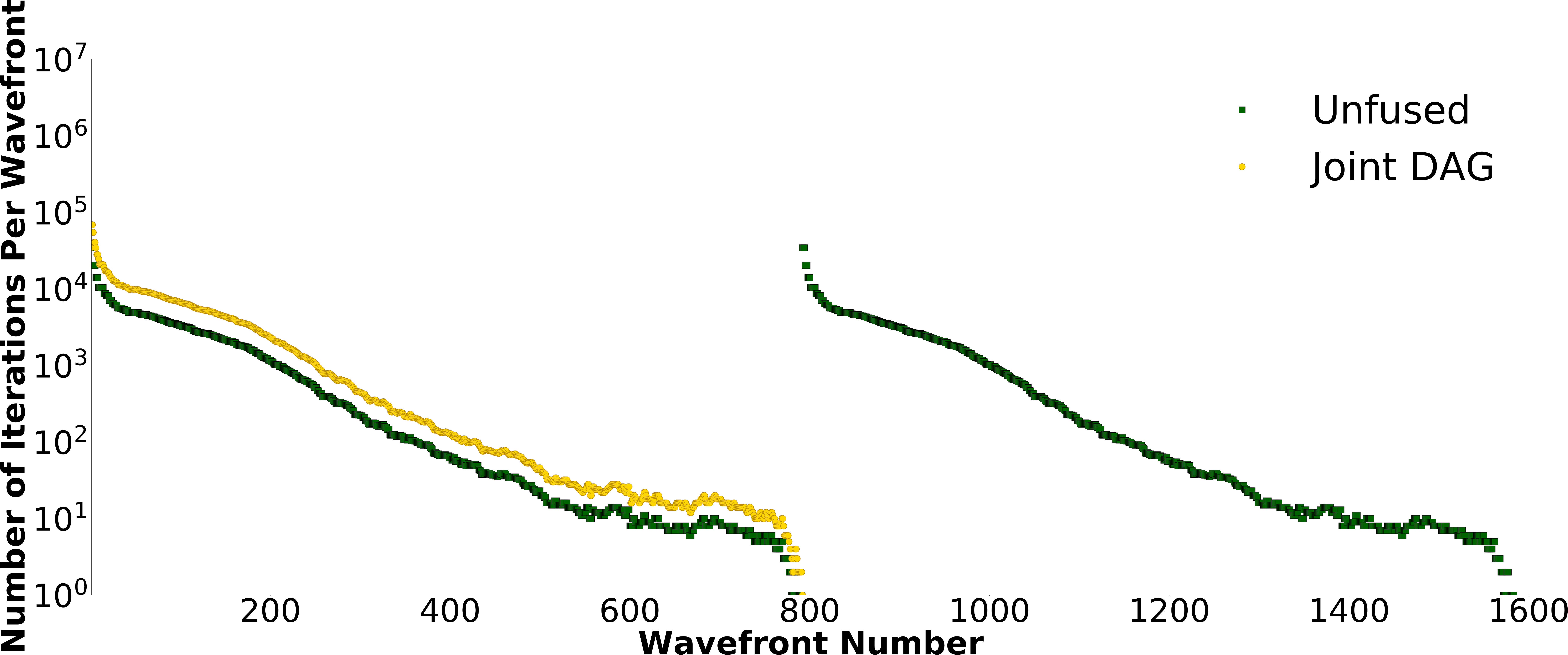}

 \caption{
The nonuniform parallelism in the DAGs of sparse incomplete Cholesky and triangular solver (annotated with unfused) and for the joint DAG of the two kernels \zedka{results in load imbalance}.  
Higher value in the y-axis shows
high parallelism in a given wavefront. Wavefront numbers
in the x-axis are numbered based on their order of execution.
 } 
 \label{fig:problem}
\end{figure}

\begin{figure*}[!ht]
\centering
\begin{tabular}{ccccc}
\centering
&&&&
  \\[-5ex]
\subfloat{
   \includegraphics[width=0.97\textwidth]{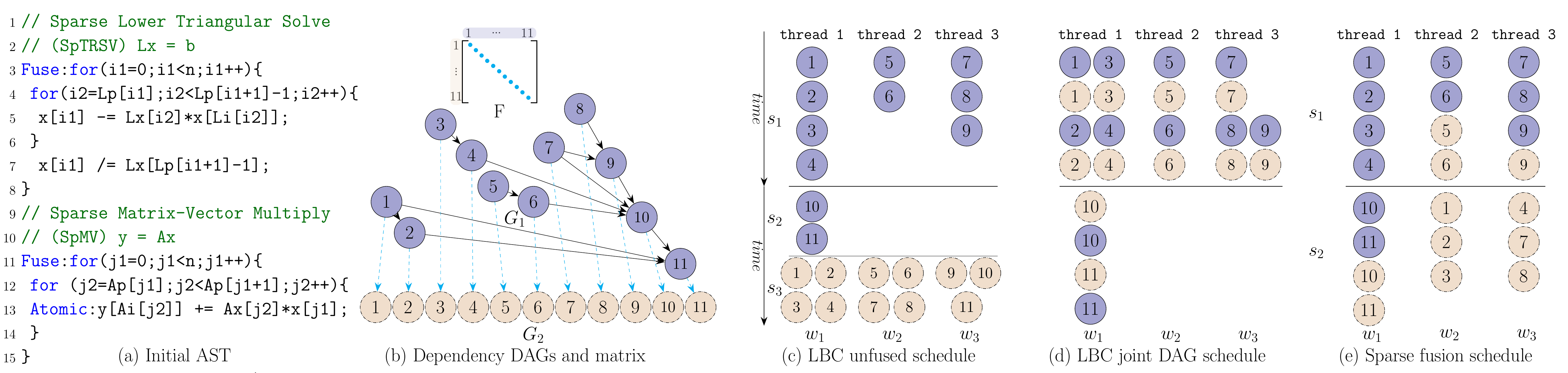}
   \label{fig:indast}
 }

  \subfloat{
\label{fig:indags}}
    &\subfloat{
\label{fig:unfused}}
 &\subfloat{
\label{fig:fusedjoint}}
&\subfloat{

\label{fig:fused}}

\end{tabular}
  \caption{ Figures~\ref{fig:unfused}-\ref{fig:fused} show three different schedules for running a sparse lower triangular kernel (SpTRSV) followed by a sparse matrix-vector multiplication (SpMV) as shown in Figure~\ref{fig:indags}. \rebt{We choose the number of processors ($r$) to be three.} \rebt{Solid} purple ($G_1$) and \rebt{dash-dotted} yellow ($G_2$) vertices in order represent iterations of SpTRSV and SpMV and edges show the dependencies between iterations. Dashed edges in Figure~\ref{fig:indags} show dependencies between two kernels \rebt{and correspond to the nonzero elements of matrix $F$}. 
  The unfused implementation schedules each DAG separately as shown in Figure~\ref{fig:unfused}. Two different fused implementations in Figure~\ref{fig:fusedjoint} and \ref{fig:fused} use  both DAGs and dependencies between kernels to build a fused schedule.
  }
\label{fig:example}
\Description[running example]{ two different implementations for the input in Listing~\ref{}.}
\end{figure*}

Numerical algorithms~\cite{saad2003iterative} and optimization methods~\cite{boyd2004convex,stellato2020osqp,cheshmi2020nasoq} 
are typically composed of numerous consecutive sparse matrix computations. 
For example, in iterative solvers~\cite{saad2003iterative} such as Krylov methods~\cite{saad1995p,chow2015fine}, sparse kernels that apply a preconditioner or update the residual are repeatedly executed inside and between
iterations of the solver. 
Sparse kernels with loop-carried
dependencies, i.e. kernels with partial parallelism, are frequently used in numerical algorithms, and the performance
of scientific simulations relies heavily on efficient parallel
implementations of these computations. 
Sparse kernels that exhibit partial parallelism often have multiple wavefronts of parallel computation where a synchronization is required for each wavefront, i.e. wavefront parallelism~\cite{venkat2016automating,govindarajan2013runtime}. The amount of parallelism varies per wavefront and often tapers off towards the end of the computation, which results in load imbalance. \zedka{Figure~\ref{fig:problem} shows with dark lines the nonuniform parallelism for the sparse incomplete Cholesky (SpIC0)  and the  sparse triangular solve (SpTRSV) kernels when SpTRSV executes after SpIC0 completes.} Separately optimizing such kernels exacerbates this problem by adding even more synchronization. Also, opportunities for data reuse between two sparse computations might not
be realized when sparse kernels are optimized separately.

\zedka{Instead of scheduling iterations of sparse kernels separately, they can be scheduled jointly. 
Wavefront parallelism can be applied to the joint DAG of two sparse computations. A data flow directed acyclic graph (DAG)  describes dependencies between iterations of a kernel~\cite{cheshmi2017sympiler,strout2018sparse,henon2002pastix}.
A joint DAG includes all of the dependencies between iterations within and across kernels. 
The joint DAG of sparse kernels with partial
parallelism with the DAG of another sparse kernel provides slightly more parallelism per wavefront without increasing the number of wavefronts. 
 The yellow line in Figure~\ref{fig:problem} shows how scheduling the joint DAG of SpIC0 and SpTRSV provides more parallelism per wavefront and significantly reduces the number of wavefronts (synchronizations).  However, the load balance issues remain, and there are still several synchronizations.}

\hkaz{Wavefront parallelism can be applied to the joint DAG of two sparse computations. A data flow directed acyclic graph (DAG)  exposes dependencies between iterations of a kernel~\cite{}. 
The joint DAG of sparse kernels with partial parallelism with the DAG of another sparse kernel typically increases available parallelism per wavefront and hence increases opportunities to improve average parallelism. However, in wavefronts with only a few computations there are still load balance issues. Figure~\ref{fig:problem} shows that the improvement in parallelism still results in poor load balance for the example kernel combination. As shown, while the joint DAG in the kernel combination provides more parallelism per wavefront without increasing the number of wavefronts, the early and late wavefronts have nonuniform parallelism.}

 \hkaz{Synchronizations of the joint DAG can be reduced further by aggregating some of the wavefronts.
DAG partitioners such as Load-Balanced Level Coarsening (LBC)~\cite{cheshmi2018parsy} and DAGP~\cite{herrmann2019multilevel} apply aggregation, however, when applied to the joint DAG because they aggregate iterations from consecutive wavefronts, load imbalance might still occur. 
For example in Figure~\ref{fig:problem}, late wavefronts of the joint DAG only have a few iterations and their aggregating them results in load imbalance. 
}

\hkaz{DAG partitioning methods potentially improve the temporal locality between the two kernels by aggregating iterations from wavefronts in the joint DAG.  But this can disturb spatial locality within each kernel. 
 For example, for two sparse kernels that only share a small array and operate on different sparse matrices, optimizing temporal locality between kernels will not be profitable.
Also,  even when applied to the DAG of an individual kernel, DAGP and LBC are slow for large DAGs because of the overheads of coarsening~\cite{}. This problem exacerbates when applied to the joint  because the joint DAG is typically 2-4$\times$ larger than an individual kernel's DAG. }

Wavefronts of the joint DAG can be aggregated to reduce the number of synchronizations. 
 DAG partitioners such as Load-Balanced Level Coarsening (LBC)~\cite{cheshmi2018parsy} and DAGP~\cite{herrmann2019multilevel} apply aggregation, however, when applied to the joint DAG because they aggregate iterations from consecutive wavefronts, load imbalance might still occur. 
 Also, by aggregating iterations from wavefronts in the joint DAG, DAG partitioning methods potentially improve the temporal locality between the two kernels but, this can disturb spatial locality within each kernel. 
 For example, for two sparse kernels that only share a small array and operate on different sparse matrices, optimizing temporal locality between kernels will not be profitable.
Finally,  even when applied to the DAG of an individual kernel, DAGP and LBC are slow for large DAGs because of the overheads of coarsening~\cite{herrmann2019multilevel}. This problem exacerbates when applied to the joint  because the joint DAG is typically 2-4$\times$  larger than an individual kernel's DAG. %

We present sparse fusion that creates an efficient schedule and fused code for when a sparse kernel with loop-carried dependencies is combined with another sparse kernel. Sparse fusion uses an inspector to apply a novel Multi-Sparse DAG Partitioning (MSP) \zedka{runtime scheduling algorithm} on the DAGs of the two input sparse kernels. MSP uses a vertex dispersion strategy to balance workloads in the fused schedule,  uses two novel iteration packing heuristics 
to improve the data locality due to spatial and temporal locality of the merged computations, and uses vertex pairing strategies to aggregate iterations without joining the DAGs.

Figure~\ref{fig:example} compares the schedule created by sparse fusion (sparse fusion schedule) with the schedules created by applying LBC to the individual DAGs  of each sparse kernels (LBC unfused schedule) and LBC applied to the joint DAG (LBC joint DAG schedule).  All  approaches take the input DAGs in Figure~\ref{fig:indags}. Solid purple vertices are the DAG of sparse triangular solve (SpTRSV) and the dash-dotted yellow  correspond to Sparse Matrix-Vector multiplication (SpMV).  LBC is a DAG partitioner that  partitions a DAG into a set of aggregated wavefronts called s-partitions that run sequentially, each s-partition is composed of some independent w-partitions.
In the LBC unfused schedule in Figure~\ref{fig:unfused}, 
 LBC is used to partition the SpTRSV DAG and  will create two s-partitions, \textit{i.e.} $s_1$ and $s_2$. The vertices of SpMV are scheduled to run in parallel in a separate wavefront  $s_3$. This implementation is not load balanced because the number of partitions that can run in parallel differs for each s-partition. 
In the LBC joint DAG schedule, the DAGs are first joint using the dependency information between the two kernels shown with blue dotted arrows and then LBC is applied to create the two s-partitions  in Figure~\ref{fig:fusedjoint}. These s-partitions  are also not load balanced, for example $s_2$ only has one partition.   Sparse fusion uses MSP to first partition the SpTRSV DAG and then disperses the SpMV iterations to create load-balanced s-partitions, e.g. the two s-partitions in  Figure~\ref{fig:fused}  have  three closely balanced partitions. 
 
 SpTRSV solves $Lx=b$ to find  $x$ and SpMV performs $y = A*x$ where $L$ is a sparse lower triangular matrix, $A$ is a sparse matrix, and $x$, $b$, and $y$ are vectors.
 The LBC joint DAG schedule interleaves iterations of two kernels to reuse \texttt{x}. However, this can disturb spatial locality within each kernel because the shared data between the two kernels, $x$, is smaller than the amount of data used within each kernel, $A$ and $L$. With the help of a reuse metric, Sparse fusion realizes the larger data accesses inside each kernel and hence packs iterations to improve  spatial locality within  each kernel.  
We implement sparse fusion as an embedded domain-specific language in \texttt{C++} that takes the specifications of the sparse kernels as input, inspects the code of the two kernels, and transforms code to  generate an efficient and correct parallel fused code.
 The primary focus of sparse fusion is to fuse two sparse kernels where at least one of the kernels has loop-carried dependence. Sparse fusion is tested on seven of the most commonly used sparse  kernel combinations in scientific codes which include kernels such as sparse triangular solver, incomplete Cholesky, incomplete LU, diagonal scaling, and matrix-vector multiplication.  The generated code is evaluated against MKL and ParSy with average speedups of  5.1$\times$ and 1.6$\times$ respectively. Sparse fusion  compared to fused implementations of LBC, DAGP, and wavefront techniques applied to the joint DAG provides on average 5.1$\times$, 7.2$\times$ and 2.5$\times$ speedup respectively.

\begin{figure}[!t] 
   \includegraphics[width=0.5\textwidth]{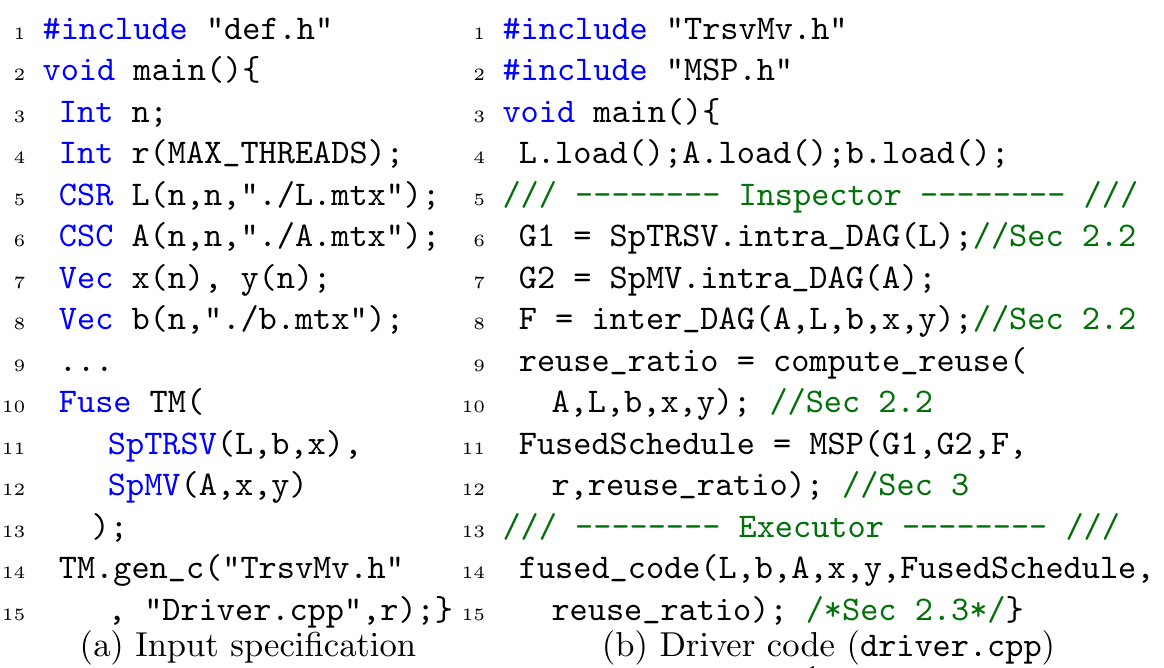}
 
 \caption{Sparse fusion's input  and  the driver code.}
 \label{fig:inputspec}
\end{figure}

\begin{figure*}[!h]
   \includegraphics[width=\textwidth]{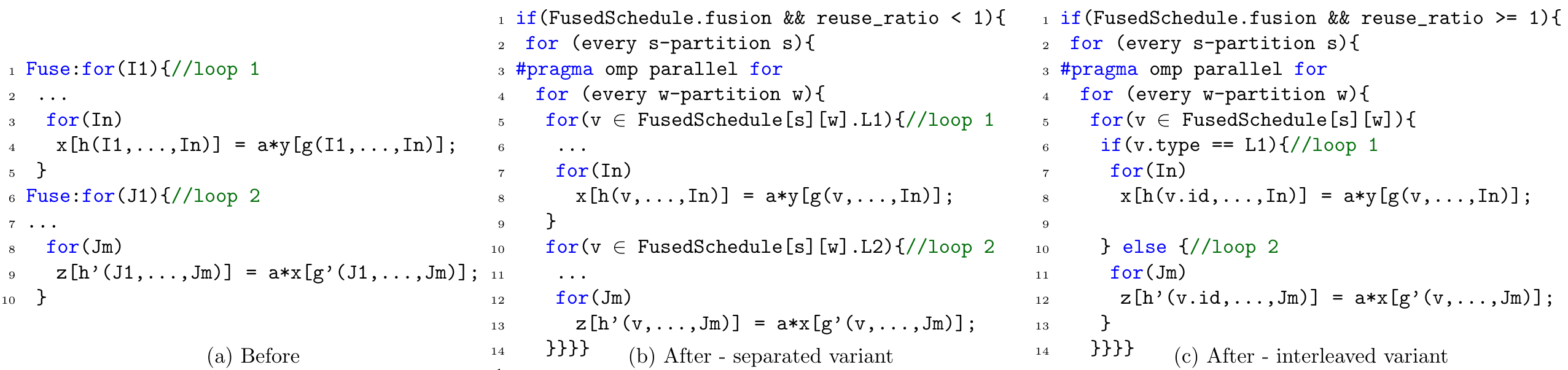}

\caption{The general form of the sparse fusion code transformation with its two variants, interleaved and separated. \rebt{
\texttt{I1...In} and \texttt{J1...Jm} represent two loop nests. \texttt{h'} and \texttt{g'} are data access functions. \texttt{FusedSchedule} contains the schedule for iterations of loops \texttt{I1}, shown with \texttt{L1} and \texttt{J1}, shown with \texttt{L2}.    
}  
}
\Description[code lowering]{the lowering template for fusion.}
\label{fig:lowering}
\end{figure*}

\section{Sparse Fusion }
 Sparse fusion is implemented  as a code generator with an inspector-executor technique that can be used as a library.  It takes the input specification shown in Figure~\ref{fig:inputspec}a and generates the inspector and the executor in Figure~\ref{fig:inputspec}b.  The inspector includes the MSP algorithm and functions that generate its inputs, i.e. dependency DAGs, reuse ratio, and the dependency matrix.    
The executor is the fused code that is created by the fused transformation.




\subsection{Code Generation}
Sparse fusion is implemented as an embedded domain-specific language. It takes as input the specification shown in Figure~\ref{fig:inputspec}a and generates the driver code in Figure~\ref{fig:inputspec}b.
At compile-time, the data types and kernels in Figure~\ref{fig:inputspec}a are converted to an initial Abstract Syntax Tree (AST) using \texttt{TM.gen\_c()} in line 14.
Lines 11 and lines 12 in Figure 3a demonstrate how the user specifies the two kernels  for the running example in Figure 2 as inputs to Sparse fusion. The corresponding AST for the example is shown in Figure 2a.


At runtime by running the driver code in Figure~\ref{fig:inputspec}b, the inspector creates a fused schedule, and the executor runs the fused schedule. 
The inspector first builds inputs to MSP using functions \texttt{intra\_DAG}, \texttt{inter\_DAG}, and \texttt{compute\_reuse} in lines 6--10 in Figure~\ref{fig:inputspec}b and then calls \texttt{MSP} in line 11 to generate \texttt{FusedSchedule} for \texttt{r} threads. Then the executor code, \texttt{fused\_code} in line 14 in Figure~\ref{fig:inputspec}b, runs in parallel using the fused schedule. 


\subsection{The Inspector in Sparse Fusion}
\label{sec:inspector}
The MSP algorithm requires kernel-specific inputs. Its inputs are  the dependency matrix between kernels, the  DAG of each kernel, a reuse ratio. Sparse fusion analyzes the kernel code, available from its AST, to generate inspector components that create these inputs. 

\textit{Dependency DAGs:} 
Lines
6--7 in Figure~\ref{fig:inputspec}b use an internal domain-specific library to generate the dependency DAG of each kernel. 
General approaches such as work by Mohammadi et al.~\cite{mohammadi2019sparse} can also be used to generate the DAGs, however, that will lead to higher inspection times compared to a domain-specific approach. 
For example, with domain knowledge, sparse fusion will  use  the $L$ matrix as the SpTRSV DAG $G_1$  in Figure~\ref{fig:indags}. Each nonzero $L_{ij}$ represents a dependency from iteration $i$ to $j$. 

\textit{Dependency Matrix $F$:}
MSP uses the dependency information between kernels to create a correct fused schedule. By running the \texttt{inter\_DAG} function, sparse fusion creates this information and stores it in matrix $F$. To generate \texttt{inter\_DAG}, sparse fusion finds dependencies between statements of the two kernels by analyzing the AST. 
Each nonzero $F_{i,j}$ represents a dependency from iteration $j$ of the first loop, i.e. column $j$ of $F$, to iteration $i$ of the second loop, i.e. row $i$ of $F$.  
In Figure~\ref{fig:indast}, there exists a read after write (flow) dependency between statements \texttt{x[i1]} in line 5 and \texttt{x[j1]} in line 13. As a result, sparse fusion generates the function shown in Listing~\ref{lst:dep_ex}. The resulting $F$ matrix, generated at runtime, is shown in Figure 2b.

\begin{lstlisting}[label={lst:dep_ex},mathescape=true,numbers=none,
caption={\texttt{inter\_DAG} function for the example in Figure~\ref{fig:indast}.}]
 for(i1=0; i1<n; i1++){
  j1 = i1;
  if(A.p[j1] < A.p[j1+1] )
   F[j1].append(i1); }
\end{lstlisting}

\textit{Reuse Ratio:}
MSP uses a reuse ratio based on the memory access patterns of the kernels  to decide whether to improve locality within each kernel or between the kernels. 
The inspector in line 9 in Figure~\ref{fig:inputspec}b computes the reuse ratio metric. The metric represents the ratio of common to total memory accesses of the two kernels, i.e. $\frac{\text{common memory access}}{max(\text{kernel1 accesses, kernel2 accesses)} }$. For a reuse ratio larger than one, the number of common memory accesses between the two kernels is larger than the accesses inside a kernel.  
Sparse fusion estimates memory accesses using the ratio of the size of common variables over the maximum of the total size of variables amongst the kernels.
For the running example, 
the code generated for \texttt{compute\_reuse}  is \texttt{2*x.n / max(A.size+x.n+y.n,L.size+\\x.n+b.n)}.  Since \texttt{x} is smaller than $L$ or $A$, the reuse ratio is less than one.


\subsection{Fused Code} 
To generate the fused code, a fused transformation is applied to the initial AST at compile-time and two variants of the fused code are generated, shown in Figure~\ref{fig:lowering}. 
The transformation variants are \textit{separated} and \textit{interleaved}. The fused code uses the reuse ratio at runtime to select the correct variant for the specific input. \zedka{The variable \texttt{fusion} in line 1 of Figure~\ref{fig:lowering}b and \ref{fig:lowering}c  is set to  \texttt{False} if MSP determines fusion is not profitable. } 
Figure~\ref{fig:lowering}a shows the  sequential loops in the  AST, which are annotated with \texttt{Fuse}, and are transformed to the separated and interleaved code variants as shown in order in Figures~\ref{fig:lowering}b and \ref{fig:lowering}c. 
%
%
The separated variant is selected when the reuse ratio is smaller than one. In this variant, iterations of one of the loops run consecutively without checking the loop type. 
The interleaved variant is chosen when the reuse ratio is larger than one. In this variant, iterations of both loops should run interleaved, and the variant checks the loop type per iteration as shown in lines 6 and 10 in Figure \ref{fig:lowering}c.

\section{\SF{} }

\label{sec:fusion}
\begin{figure*}[!ht]
\centering
\renewcommand\arraystretch{0}
\begin{tabular}{cccc}
\subfloat{
\label{fig:step11}}
 &\subfloat{
\label{fig:step12}}
&\subfloat{
\label{fig:step21}}
&\subfloat{
\label{fig:step22}}
\\[-6ex]
\subfloat{
   \includegraphics[width=0.97\textwidth]{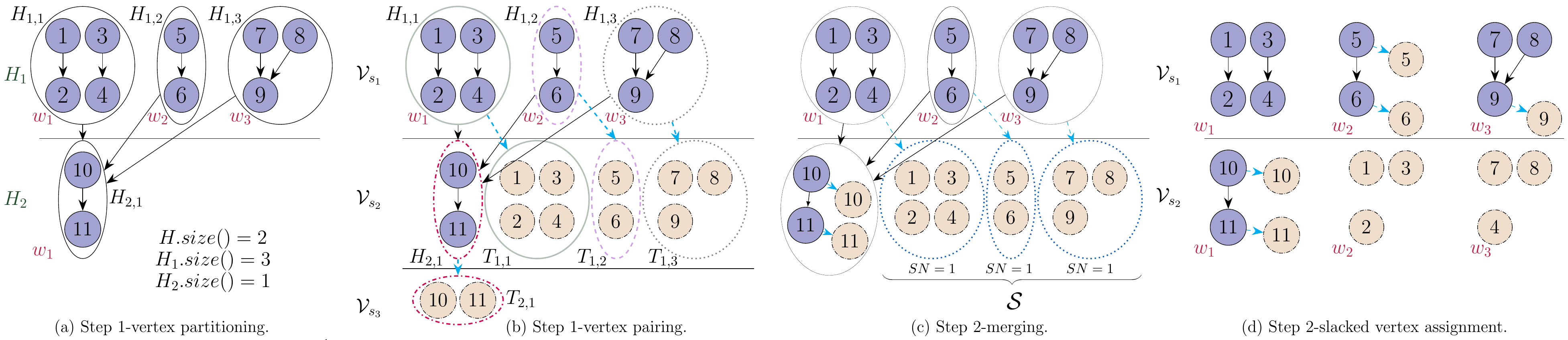}
 }

\end{tabular}
  \caption{
  Stages of  \algo{}  for DAGs $G_1$ and $G_2$ \rebt{ and matrix $F$ in the}  running example shown in Figure~\ref{fig:indags} \rebt{where the reuse ratio ($reuse\_ratio$) is smaller than one and number of processors ($r$) is three}. The first step of the algorithm selects  $G_1$ and creates $H$ partitioning \rebt{for three processors} using the LBC algorithm as shown in Figure~\ref{fig:step11}. Then it pairs each $H_{i,j}$ through dependencies in matrix $F$ to create partitioning $T$ of $G_2$ as shown in Figure~\ref{fig:step12}. The partitions with the same line pattern/color are pair partitions. In the second step, \algo{} merges pair partitions that cannot be dispersed such as first w-partitions of s-partitions 2 and 3 ($\mathcal{V}_{s_3,w_1}$ and $\mathcal{V}_{s_2,w_1}$) in Figure~\ref{fig:step12}, these are merged into $\mathcal{V}_{s_2,w_1}$ in Figure~\ref{fig:step21}. Slack vertices,  \rebt{which are denoted as $\mathcal{S}$} are shown with blue dotted circles in Figure~\ref{fig:step21}. Slack vertices are assigned into imbalanced w-partitions  as shown in Figure~\ref{fig:step22}. \rebt{Since the reuse ratio is smaller than one, vertices inside each partition are packed separately as shown}  \rebtout{ after reordering is} in Figure~\ref{fig:fused}. 
  }
\label{fig:alg_ex}
\Description[running example]{ two different implementations for the input in Listing~\ref{}.}
\end{figure*}

Sparse fusion uses  the \sfu{} (\algo{}) algorithm to create 
an efficient fused partitioning that will be used to schedule iterations of the fused code. 
\algo{} partitions vertices of the DAGs of the two input kernels to create parallel load-balanced workloads for all cores while improving locality within each thread. 
\zedka{This section describes the inputs, output, and three steps of the MSP algorithm using the running \textbf{example} in Figures~\ref{fig:example} and ~\ref{fig:alg_ex}. }
\hkaz{To maintain a good load balance and to improve locality,  \algo{}  goes through three steps. The first step creates a fused partitioning with independent workloads to run in parallel on the processor cores. The second step balances the workloads and reduces the number of synchronizations. Finally, locality in each thread is improved by creating an efficient order of execution for the workloads. }

\begin{algorithm}[!ht]
\SetAlgoLined
\DontPrintSemicolon
\begin{small}
\caption{\label{alg:datafusionAlg}\rebt{The \algo{} algorithm.} 
}
\SetKwInOut{Input}{Input}
\SetKwInOut{Output}{Output}
\Input{\,$G_1(V_1,E_1,c_1)$, $G_2(V_2,E_2,c_2)$, $F$, $r$, $reuse\_ratio$}
\Output{\,$\mathcal{V}$}
\tcc{(i) Vertex partitioning and partition pairing}
  \eIf{$|E_2| > 0 $}{\label{lin:alg1-pair1}
  
      $[H,k]$ = LBC($G_2$, $r$).list(), $T=\emptyset$, $\mathcal{V} = \emptyset $\; \label{lin:alg1-p11}
  \tcc{Backward pairing}
      \For{$ (i=1 : H.size() )$ }{\label{lin:alg1-pairs11}
      \For{$ (j=1 : H_i.size())  $ }{\label{lin:alg1-pairs111}
          $T_{i,j} =$ BFS($H_{i,j}, F, G_1$)\; \label{lin:alg1-sc1}
          $\mathcal{V}.add(T_{i,j}, H_{i,j})$\;\label{lin:alg1-sc1-1} 
      }
      }\label{lin:alg1-pairs12}
   \lIf{$ |\mathcal{V}| > 2\times(|V_1| + |V_2|)$ }{
        \
        $\mathcal{V}$.fusion =  False, exit() \label{lin:const-red}
    }
  } {\label{lin:alg1-pair3} \label{lin:alg1-pair2}
    $[H,k]$ = LBC($G_1$, $r$).list(), $T=\emptyset$, $\mathcal{V} = \emptyset $ \;\label{lin:alg1-p12}
    \tcc{Forward pairing}
    \For{$ (i=1 : H.size() )$ }{\label{lin:alg1-pairs21}
      \For{$ (j=1 : H_i.size())  $ }{\label{lin:alg1-pairs211}
         $T_{i,j} =$ BFS($H_{i,j}, F^T, G_2$)\; \label{lin:alg1-sc2-1}

         $U_{i,j}= T_{i,j}$.remove\_uncontained($F$)\; \label{lin:alg1-sc2}
        $\mathcal{V}.add(H_{i,j}, T_{i,j}, U_{i,j})$\;\label{lin:alg1-sc2-3}
      }
    }\label{lin:alg1-pairs22}
}

\tcc{(ii) Merging and slacked vertex assignment } 
$\mathcal{S}$ = slack\_info($\mathcal{V}$)\;\label{lin:alg1-s1}
\For{(every w-partition pair $(w, w') \in \mathcal{V}.pairs )$}{ \label{lin:alg1-s11}
    \lIf{$(SN(w) = 0) \land (SN(w') = 0)$}{
      $\mathcal{V}$.merge($w$,$w'$) \label{lin:alg1-s11-1}
    }
}\label{lin:alg1-s12}

$\mathcal{V} = \mathcal{V} - \mathcal{S}$,  $\epsilon = |\mathcal{V}| \times 0.001$\; \label{lin:alg1-s2}
\For{$ (i=1 : \mathcal{V}.b)$ }{\label{lin:alg1-sv1}
      \For{$ (j=1 : m_i  )$ }{
    \lIf{max\_diff($\mathcal{V}_{s_i},\mathcal{V}_{s_i,w_j}$) > $\epsilon \land \mathcal{S} \neq \emptyset$ }{ \label{lin:maxdiff}
        $\mathcal{S} = \mathcal{V}_{s_i,w_j}$.balance\_with\_pair($\mathcal{S}$) \label{lin:alg1-s3}
    }
    \lIf{max\_diff($\mathcal{V}_{s_i},\mathcal{V}_{s_i,w_j}$) > $\epsilon \land \mathcal{S} \neq \emptyset$ }{\label{lin:maxdiff2}
        $\mathcal{S} = \mathcal{V}_{s_i,w_j}$.balance\_with\_slacks($\mathcal{S}$) \label{lin:alg1-s4}
 }

}
 \lIf{$\mathcal{S} \neq \emptyset$} {   
      $\mathcal{S} = \mathcal{V}_{s_i}$.assign\_even($\mathcal{S}$)\label{lin:alg1-s4-1}
  }
}\label{lin:alg1-s5}
\tcc{(iii) Packing}
\lIf{$reuse\_ratio \geq 1$}{\label{lin:alg1-p1}
    $\mathcal{V}$.interleaved\_pack($F$)\    
}\lElse{
    $\mathcal{V}$.separated\_pack()\ \label{lin:alg1-p2}
}

\end{small}

\end{algorithm}

\subsection{Inputs and Output to MSP}
 \hkaz{\algo{}  takes two DAGs annotated with a cost model, an inter-DAG dependency matrix, the number of requested partitions, and the reuse ratio as inputs and generates a fused partitioning that will schedule iterations of the fused code in parallel.}

The inputs to MSP (shown in Algorithm~\ref{alg:datafusionAlg}) are two DAGs $G_1$ and $G_2$ from in order lexicographically first and second input kernels, and the inter-DAG dependency matrix $F$ that stores the dependencies between kernels. A DAG  shown with $G_j(V_j,E_j,c)$ has a vertex set $V_j$ and an edge set $E_j$ and a non-negative integer weight $c(v_i)$ for each vertex $v_i \in V_j$. 
The vertex $v_i$ of $G_j$ represents iteration $i$ of a kernel and each edge  shows a dependency between two iterations of a kernel. $c(v_i)$ is the computational load of a vertex and is defined as the total number of nonzeros touched to complete its computation. Because sparse matrix computations are generally memory bandwidth-bound, $c(v_i)$ is a good metric to evaluate load balance in the algorithm~\cite{cheshmi2018parsy}.
\hkaz{Each nonzero $F_{i,j}$ in matrix $F$ represents a dependency from $v_j$ of  $G_1$ to $v_i$ of  $G_2$. }
$F$ is stored in the compressed sparse row (CSR) format and $F_i$ is used to extract the set of vertices in $G_1$ that $v_i \in V_2$  depends on. 
Other inputs to the algorithm are the number of requested partitions $r$, which is set to the number of cores, and the reuse ratio discussed in section~\ref{sec:inspector}.

The output of \algo{} is a \textit{fused partitioning} $\mathcal{V}$ that has $b\geq1$ s-partitions, each s-partition contains  up to $k>1$ w-partitions, where $k \le r$.
\algo{} creates $b$ disjoint s-partitions from vertices of both DAGs, shown with  $\mathcal{V}_{s_i}$ where $\cup_{i=0}^{b} \mathcal{V}_{s_i} = V_1 \cup V_2$. Each s-partition includes vertices from a lower bound and upper bound of wavefront numbers shown with $s_i = [lb_i .. ub_i)$ as well as some \textit{slack vertices}.
For each s-partition $\mathcal{V}_{s_i}$, \algo{} creates $m_i \leq k$ independent \emph{w-partitions}  $\mathcal{V}_{s_i, w_j}$ where    $\mathcal{V}_{s_i,w_1}\cup...\cup \mathcal{V}_{s_i,w_{m_i}}=\mathcal{V}_{s_i}$.
Since w-partitions are independent, they can run in parallel. 

\textbf{\textit{Example.}} 
\zedka{In Figure~\ref{fig:indags}, the SpTRSV DAG $G_1$, the SpMV DAG $G_2$, the inter-DAG dependency matrix $F$ are inputs to MSP. Other inputs to MSP are $r$=3 and the $reuse\_ratio$.}
The fused partitioning shown in Figure~\ref{fig:fused} has two s-partitions ($b$=2). The first s-partition  has three w-partitions ($m_1$=3) shown with $\mathcal{V}_{s_1} =  \{ [\underline{1}, \underline{2}, \underline{3}, \underline{4} ]; [\underline{5},\underline{6}, 5, 6 ]; [\underline{7},\underline{8},\underline{9}, 9]  \}$,  the underscored vertices belong to  $G_1$.

\subsection{The \algo{} Algorithm}
Algorithm~\ref{alg:datafusionAlg} shows the \algo{} algorithm. It takes the inputs and goes through three steps of \textit{(1)} vertex partitioning and partition pairing with the objective to aggregate iterations without joining the DAGs of the inputs kernels; \textit{(2)} merging and slack vertex assignment to reduce synchronization and to balance workloads; and \textit{(3)} packing to improve locality. 

\subsubsection {Vertex Partitioning and Partition Pairing.}
The first step of  MSP  partitions one of the input DAGs $G_1$ or $G_2$, and then uses that partitioning to partition the other DAG. The created partitions are stored in  $\mathcal{V}$. 
Partitioning the joint DAG 
\hkaz{the combined DAG of both kernels with their inter-DAG dependencies,}
is complex and might not be efficient because of the significantly larger number of edges and vertices added compared
to the individual DAG of each kernel. Instead, MSP ignores the dependencies across kernels and first creates a partitioning from one of the DAGs with the help of \textit{vertex partitioning}. Then the other DAG  is partitioned using a \textit{partition pairing} strategy.  The DAG that is partitioned first is the head DAG and the other is the tail
DAG. 
\hkaz{Since dependencies between iterations imply data is shared between them, putting dependent iterations in the same partition improves locality. Therefore, MSP always chooses the DAG of the kernel with dependencies as the head DAG. When both $G_1$ and $G_2$ have edges, the head DAG will be $G_2$ to reduce the inspector overhead.  The final partitions are  assigned to s-partitions and w-partitions and are stored in the fused partitioning  $\mathcal{V}$.  }
\zedka{\textit{A head DAG choice strategy} is used to select the head DAG.}

\emph{Vertex partitioning}. MSP uses the LBC DAG partitioner~\cite{cheshmi2018parsy} \hkaz{to partition one of the DAGs which we call the head DAG and}to construct a partitioning of the head DAG \zedka{in  lines~\ref{lin:alg1-p11} and \ref{lin:alg1-p12} of Algorithm~\ref{alg:datafusionAlg} by calling the function \texttt{LBC}}.
The resulting partitioning has a set of disjoint s-partitions. Each s-partition contains $k$  disjoint w-partitions which are balanced using vertex weights. Disjoint w-partitions ensure all w-partitions within s-partitions are independent.  
\hkaz{In  lines~\ref{lin:alg1-p11} and \ref{lin:alg1-p12} of Algorithm~\ref{alg:datafusionAlg},  function \texttt{LBC} \hkaz{takes the head DAG and the number of requested partitions $r$ as inputs and} partitions the head DAG.} The created partitions are  stored in a two-dimensional list $H$ using \texttt{list}, e.g. w-partition $w_j$ of s-partition $s_i$ is stored in $H_{ij}$. 

\emph{Partition pairing}. The algorithm then partitions the tail
DAG with \textit{forward pairing}, if $G_1$ is the head DAG, or with  \textit{backward pairing}, if $G_2$ is the head DAG. With the pairing strategy, 
some of the partitions of the tail DAG are paired with the head DAG partitions.
Pair-partitions are \textit{self-contained} so that they execute in parallel if assigned to the same s-partition. 
The created partitions are put in the fused partitioning  $\mathcal{V}$ to be used in step two. The following first describes the condition for partitions to be self-contained and then explains the forward and backward pairing strategies.

Pair partitions $H_{ij}$ and $T_{ij}$ are called self-contained if all reachable vertices from a breadth first search (BFS) on $\forall v \in H_{ij} \cup T_{ij}$ through vertices of $G_1$ and $G_2$ are in $H_{ij} \cup T_{ij}$. 
Self-contained pair partition $(H_{ip}, T_{ip})$ and pair partition $(H_{iq}, T_{iq})$ can execute in parallel without  synchronization if  in the same wavefront $i$, 
i.e.  $\forall 1 \leq i \leq b \land ( 1 \leq p,q \leq m_i) $. Partitions that do not satisfy this condition create synchronizations in the final schedule.

\hkaz{Assuming $G_1$ and $G_2$ are in order the DAG of lexicographically first and  second kernels, backward pairing } \zedka{The backward pairing strategy} visits every partition $H_{i,j}$ and performs a BFS (line~\ref{lin:alg1-sc1}) from vertex $v_l \in H_{i,j}$ to its dependent vertices in $G_1$ which are reachable through $F_l$. 
Reachable vertices are stored in $T_{ij}$. 
The partitions in $H$ and $T$ are assigned a w- and s-partition and are then put in the fused partitioning $\mathcal{V}$ (\zedka{via \texttt{add}} in line~\ref{lin:alg1-sc1-1}).   The assigned s- and w-partitions for $H_{ij}$ are $s_{i+1}$ and $w_j$ respectively, i.e. $\mathcal{V}_{s_{i+1},w_{j}}$. 
$T_{ij}$ should be executed before $H_{ij}$ thus is placed in s-partition $s_{i}$ or $\mathcal{V}_{s_{i},w_{m_i+1}}$, where $m_i$ is number of w-partitions in $\mathcal{V}_{s_i}$ at this point.  If a vertex in $H_{i,j}$ depends on more than one vertex in $G_1$, some vertices are replicated in different $T$ partitions. While replication leads to redundant computation, it ensures that the pair partition $(H_{i,j},T_{i,j})$ is self-contained because vertices that depend on the vertices in  $H_{i,j}$ will be included in $T_{i,j}$. 
\rebt{MSP performs fusion only if profitable, hence fusion is disabled (by setting \texttt{fusion} to \texttt{False}) if the number of redundant computations go beyond a threshold. This threshold is $2\times(|V_1| + |V_2|)$ in line~\ref{lin:const-red} and is defined as the sum of vertices of both DAGs. } 

The \zedka{forward} pairing strategy iterates over every partition $H_{i,j}$ and performs a BFS from vertex $v_l \in H_{i,j}$ to its reachable vertices in $G_2$ through $F^T_l$, see lines~\ref{lin:alg1-pairs21}--\ref{lin:alg1-pairs22} in Algorithm~\ref{alg:datafusionAlg}. The list of reachable vertices are stored in $T_{i,j}$ \hkaz{as shown}\zedka{via \texttt{BFS}} in line~\ref{lin:alg1-sc2-1}.  
If a vertex $v_m$ in $T_{i,j}$ depends on vertex $v_l$ in $G_1$ and $v_l$ does not exist in $H_{i,j}$ then $v_m$ should be removed to ensure $(H_{i,j},T_{i,j})$ is self contained. \zedka{The \texttt{remove\_uncontained} function in line~\ref{lin:alg1-sc2} removes vertex $v_m$ and puts it in  partition $U_{i,j}$.} \hkaz{The removed vertex $v_m$ is put in  partition $U_{i,j}$ \rebt{(line~\ref{lin:alg1-sc2} of Algorithm~\ref{alg:datafusionAlg})}.}  
 Finally, the created partitions are assigned to the fused partitioning $\mathcal{V}$ \zedka{via \texttt{add} in line~\ref{lin:alg1-sc2-3}} as follows: $\mathcal{V}_{s_i,w_{j}} = H_{i,j}$,
 $\mathcal{V}_{s_{i+1},w_{m_{i+1}+1}} = T_{i,j} $, $\mathcal{V}_{s_{i+1},w_{m_{i+1}+1}} = U_{i,j}$.

\hkaz{Lines~\ref{lin:alg1-pairs11} and \ref{lin:alg1-pairs111} in Algorithm~\ref{alg:datafusionAlg} perform backward pairing by first  iterating over every s-partition $i$ between 1 to $H.size()$ and every w-partition $j$ inside the s-partition $i$ which is between 1 to $H_i.size$. Then in line~\ref{lin:alg1-sc1},  \texttt{BFS} is applied to all vertices in $H_{i,j}$ and the reached vertices are stored in $T_{i,j}$. The function \texttt{add} in line~\ref{lin:alg1-sc1-1} stores both $H_{i,j}$ and $T_{i,j}$ in the fused partitioning $\mathcal{V}$. For forward pairing, the additional step of removing uncontained vertices is performed in line~\ref{lin:alg1-sc2} by calling the \texttt{remove\_uncontained} function. }

\textit{The head DAG choice}. MSP chooses the DAG with edges as the head DAG to improve locality.  Locality is improved because the head DAG is partitioned with LBC. LBC creates well-balanced partitions with good locality when applied to DAGs with edges. Selecting $G_2$ as the head DAG reduces inspector overhead.
If both $G_1$ and $G_2$ are DAGs of kernels with dependency, then $G_2$ is chosen as the head DAG to reduce inspector overhead. When $G_2$ is partitioned first, MSP chooses backward pairing which is more efficient compared to forward pairing.  \rebt{Forward pairing   traverses $F$ and its transpose $F^T$ and thus performs $2*nnz_F + 2*n$ operations where $nnz_F$ is the number of nonzeros in $F$. However, backward pairing only traverses $F$ and performs $nnz_F + n$ operations}.   


\textit{\textbf{\textit{Example.}} } 
\zedka{Figures~\ref{fig:step12} shows the output of MSP after the first step for the inputs in Figure~\ref{fig:indags}.}
MSP chooses $G_1$ as the head DAG because it has edges ($|E_1|>1$), $G_2$ has no edges.   In vertex partitioning, $G_1$ is partitioned with LBC to create up to three w-partitions (because $r=3$) per s-partition. The created partitions are shown in Figure~\ref{fig:step11} and are stored in $H$. The first s-partition $\mathcal{V}_{s_1}$ is stored in $H_1$ and its three w-partitions are indexed with $H_{1,1}$, $H_{1,2}$, and $H_{1,3}$. Similarly, $\mathcal{V}_{s_2}$ is stored $H_2$ and its only w-partition is in $H_{2,1}$. 
Figure~\ref{fig:step12} shows the output of partition pairing.
Since $G_1$ is the head DAG,  MSP uses forward pairing and performs a BFS from each partition in  $H$ to create self-contained pair partitions stored in $T$.
For example, a BFS from $H_{1,1}=\{\underline{1},\underline{2},\underline{3},\underline{4}\}$ creates $T_{1,1}=\{1,2,3,4\}$. Since $T_{1,1}$ and $H_{1,1}$ are self-contained, no vertices are removed from  $T_{1,1}$ and thus $U_{1,1}=\emptyset$.
Finally, MSP puts $H_{1,1}$ and $T_{1,1}$ in $\mathcal{V}_{s_1,w_1}$ and $\mathcal{V}_{s_2,w_2}$ respectively, and adds $(\mathcal{V}_{s_1,w_1},\mathcal{V}_{s_2,w_2})$ to   $\mathcal{V}.pairs$. The final partitions and pairings  as shown in Figure~\ref{fig:step12} are:
 $\mathcal{V} =[ \{  H_{1,1}, H_{1,2}, H_{1,3}  \}, \{ H_{2,1}, T_{1,1}, T_{1,2}, T_{1,3}  \}, \{ T_{2,1} \} ]  =  [\{ \{\underline{1},\underline{2}, \underline{3}, \underline{4}\},\\ \{\underline{5}, \underline{6} \}, \{\underline{7}, \underline{8}, \underline{9}\} \},  \{ \{\underline{10},\underline{11}\}, \{1, 2, 3, 4 \}, \{5,6\}, \{7, 8, 9\} \},  \{ \{10, 11\}\\  \}  ]$
  and the pairing information is:
 $ \mathcal{V}.pairs = \{(\mathcal{V}_{s_1,w_1},\mathcal{V}_{s_2,w_2}),\\ (\mathcal{V}_{s_1,w_2},\mathcal{V}_{s_2,w_3}),
 (\mathcal{V}_{s_1,w_3},\mathcal{V}_{s_2,w_4}),
 (\mathcal{V}_{s_2,w_1},\mathcal{V}_{s_3,w_1})\}$.

\subsubsection{Merging and Slack Vertex Assignment.} \zedka{The second step of  \algo{} reduces the number of synchronizations  by merging some of the pair partitions in a \textit{merging} phase. It also improves load balance by dispersing vertices across partitions using \textit{slacked vertex assignment}. }  
\hkaz{The second step of  \algo{} reduces the number of synchronizations  by merging some of the pair partitions  and improves load balance by dispersing some vertices across partitions.  
 Although the partitioning from step one, aims to create enough parallel workloads, it might not be successful because of existing dependencies between partitions. To resolve this, we merge  partition pairs that should execute  only in their assigned s-partitions. Merging partitions typically reduces the number of synchronizations and hence improves parallelism.
 Also even with enough parallel workloads for all cores, the partitions might be imbalanced. 
To improve load balance, this step disperses across partitions, vertices that their execution can be postponed. To determine which partitions to merge and which vertices to disperse, \algo{} finds slack vertices and their corresponding slack number. In the following, we provide the required slack definitions and explain how partition pairs are merged and vertices are dispersed. }

\emph{Slack definitions:}  
%
%
A vertex $v$ can always run in its wavefront number $l(v)$. However, the execution of vertex $v$ can sometimes  be postponed up to $SN(v)$ wavefronts  without having to move its dependent vertices to later wavefronts. 
$SN(v)$ is the slack number of $v$ and is defined as
 $SN(v) = P_G - l(v) - height(v)$ where $height(v)$ is the maximum path from a vertex $v$ to a sink vertex (a sink vertex is a vertex without any outgoing edge), $P_G$ is the critical path of $G$, and $l(v)$ is the wavefront number of $v$. A vertex with a positive slack number is a \emph{slack vertex}.
To compute vertex slack numbers efficiently, instead of visiting all vertices,  \algo{} iterates over partitions and computes the slack number of each partition in the partitioned DAG, i.e. \emph{partition slack number}. The computed slack number for a partition is assigned to all vertices of the partition. 
As shown in line~\ref{lin:alg1-s1} of Algorithm~\ref{alg:datafusionAlg}, all partition slack numbers of $\mathcal{V}$ are computed \zedka{via \texttt{slack\_info}} 
and are stored in $\mathcal{S}$. 
For example, because vertices in $\mathcal{V}_{s_2,w_3}$ can be postponed one wavefront, from s-partition 2 to 3, their slack number is 1. Vertices in w-partitions $\mathcal{V}_{s_2,w_1}$ and $\mathcal{V}_{s_3,w_1}$  can not be moved because their slack numbers are zero.

\emph{Merging.} \zedka{MSP finds pair partitions with partition slack number of zero and then merges them as shown in lines~\ref{lin:alg1-s11}-\ref{lin:alg1-s12}.} \hkaz{in \algo{} finds pair partitions with partition slack number of zero, and then merges them to reduce the number of synchronizations and  improve parallelism (lines~\ref{lin:alg1-s11}-\ref{lin:alg1-s11-1} of Algorithm~\ref{alg:datafusionAlg}). } 
Since  pair partitions are self contained, merging them does not affect the correctness of the schedule.  
Algorithm~\ref{alg:datafusionAlg} visits all pair partitions $(w,w')$ in $\mathcal{V}.pairs$ and merges them using the \texttt{merge} function in line~\ref{lin:alg1-s11-1}  if their slack numbers are zero, i.e. $SN(w)=0$ and $SN(w')=0$.  
The resulting merged partition is stored in $\mathcal{V}$ in place of the w-partition with the smaller s-partition number.  
\hkaz{In the example, the algorithm merges pair partitions  $(\mathcal{V}_{s_2,w_1}, \mathcal{V}_{s_3,w_1})$, shown with red dash-dotted circles in Figure~\ref{fig:step12} because their slack numbers are zero. And then it places the merged partition in $\mathcal{V}_{s_2,w_1}$ to reduce one synchronization as shown in Figure~\ref{fig:step21}. }

\textit{Slacked vertex assignment.} \zedka{The algorithm then uses slacked vertex assignment}\hkaz{is used} to approximately load balance the w-partitions of an s-partition using a cost model. \hkaz{Slacked vertices are moved across different s-partitions and w-partitions.} The cost of  w-partition $w \in \mathcal{V}_{s_i}$ is defined as $cost(w) = \sum_{v\in w}c(v)$.
A w-partition  is balanced if the maximal difference of its cost and the cost of other w-partitions in its s-partition is smaller than a  threshold $\epsilon$.
\zedka{The maximal difference for a w-partition inside a s-partition is computed by subtracting its cost from the cost of the w-partition (from the same s-partition) with the maximum cost.}

MSP  first removes all slacked vertices $\mathcal{S}$ from the fused partitioning $\mathcal{V}$ in line~\ref{lin:alg1-s2}. 
It then goes over every s-partition $i$ and w-partition $j$ and balances  $\mathcal{V}_{s_i,w_j}$ by assigning a slacked vertex to it where possible.
W-partition $\mathcal{V}_{s_i,w_j}$ becomes balanced with vertices from its pair partition using the function \texttt{balance\_with\_pair} in line~\ref{lin:alg1-s3}.
If $\mathcal{V}_{s_i,w_j}$ is still imbalanced, \texttt{balance\_with\_slacks} in line~\ref{lin:alg1-s4} balances the w-partition using the slacked vertices $v_l\in\mathcal{S}$ that satisfy the following condition $l(v_l)<i<(l(v_l)+SN(v_l))$. 
Slack vertices in $\mathcal{S}$ that depend on each other are dispersed as a group to the same w-partition for correctness.
In line~\ref{lin:alg1-s4-1}, slacked vertices in $\mathcal{S}$ that are not postponed to later s-partitions are evenly divided between the w-partitions of the current s-partition ($\mathcal{V}_{s_i}$) using the \texttt{assign\_even} function.

\hkaz{\rebtout{The algorithm first puts all slacked vertices into a set, shown with $S$}\rebt{The MSP algorithm first removes all computed slacked vertices $\mathcal{S}$ from the fused partitioning $\mathcal{V}$,} in line~\ref{lin:alg1-s2} of Algorithm~\ref{alg:datafusionAlg}. Then in lines 25-30 it goes over each s-partition and balances every imbalanced w-partition by assigning a slacked vertex where possible to keep the maximal difference of the w-partition below a the threshold $\epsilon$.
Slacked vertex $v_l\in\mathcal{S}$ is either assigned to its pair partition or to a w-partition in $\mathcal{V}_{s_i}$ that the vertex slack range $[l(v_l) ... (l(v_l) + SN(v_l))]$ has an overlap with $s_i$. At the end, the w-partitions contain vertices from one or both DAGs and are balanced to a threshold. 
\rebt{Lines 25 and 26, iterate over all s-partitions,  shown with $\mathcal{V}.b$, and the  $m_i$ w-partitions inside each s-partition $i$.  The maximal difference between $\mathcal{V}_{s_i,w_j}$ and other w-partitions in $\mathcal{V}_{s_i}$ is found via the \texttt{max\_diff($\mathcal{V}_{s_i}$, $\mathcal{V}_{s_i,w_j}$)} function in lines~\ref{lin:maxdiff} and \ref{lin:maxdiff2}. Slacked vertices are assigned to pair partitions or to w-partitions using the functions \texttt{balance\_with\_pair} and \texttt{balance\_with\_slacks} in the algorithm in order.  \rebt{In line~\ref{lin:alg1-s4-1}, slacked vertices in $\mathcal{S}$ that can not be slacked to later levels are evenly divided between the w-partitions of the current s-partition ($\mathcal{V}_{s_i}$) using the \texttt{assign\_even} function.  }
}
}

\textit{\textbf{Example.}}
\zedka{Figure~\ref{fig:step22} shows the output of the second step of MSP from the partitioning in Figure~\ref{fig:step12}.}
First pair partitions  $(\mathcal{V}_{s_2,w_1}, \mathcal{V}_{s_3,w_1})$, shown with red dash-dotted circles in Figure~\ref{fig:step12}, are merged because their slack numbers are zero. The resulting merged partition is placed in $\mathcal{V}_{s_2,w_1}$ to reduce synchronization as shown in Figure~\ref{fig:step21}.
Then slacked vertex assignment balances the w-partitions in Figure~\ref{fig:step21}. The balanced partitions are shown in Figure~\ref{fig:step22}.
The slacked vertices $S$, are shown with dotted blue circles in Figure~\ref{fig:step21}. 
The w-partitions in $\mathcal{V}_{s_1}$  are balanced  using vertices of their pair partitions, e.g. the yellow dash-dotted vertices 5 and 6 are moved to $w_2$ in $\mathcal{V}_{s_1}$ as shown in Figure~\ref{fig:step22}. \texttt{balance\_with\_slacks} is used to balance partitions in $\mathcal{V}_{s_2}$. This is because the vertices in $S$ do not belong to the pair partitions of the w-partitions in $\mathcal{V}_{s_2}$. However, since the  slack vertices in $S$ can execute in either s-partition two or three because they are from s-partition one and have a slack number of one, they are  used to balance the w-partitions in $\mathcal{V}_{s_2}$.

\subsubsection{Packing.}
The third step  of \algo{} reorders the vertices inside a w-partition to improve data locality for a thread within each kernel or between the two kernels. 
The previous steps of the algorithm create w-partitions that are  composed of vertices of one or both kernels however the order of execution is not defined.
Using the reuse ratio, \hkaz{provided as input by the \texttt{compute\_reuse} function in the inspector (Figure~\ref{fig:inputspec}b) to the algorithm,}
the order at which the nodes in a w-partition should be executed is determined with a packing strategy.
MSP has two packing strategies: \textit{(i)} in interleaved packing,  the  vertices of the two DAGs in a w-partition are interleaved for execution and \textit{(ii)} in separated packing the vertices of each kernel are executed separately.   
Interleaved packing improves temporal locality between kernels while separated packing enhances spatial and temporal locality within kernels.
When the reuse ratio is greater than one, in line~\ref{lin:alg1-p1} of Algorithm 1 function \texttt{interleaved\_pack} is called to interleave iterations of the two kernels based on F. Otherwise,  \texttt{separated\_pack} is called (line~\ref{lin:alg1-p2}) to pack iterations of each kernel separately. 

\textit{\textbf{Example.}}
\zedka{Figure~\ref{fig:fused} shows the output of  MSP's third step from the partitioning in Figure~\ref{fig:step22}.}
\zedka{Since the reuse ratio is smaller than one separated packing is chosen thus}
\hkaz{For the fused partitioning of the example shown in Figure~\ref{fig:fused} the reuse ratio is smaller than one, thus separated packing is chosen and}$\mathcal{V}_{s_2,w_1}$ is stored as $\mathcal{V}_{s_2,w_1}=\{ [\underline{10},\underline{11}, 10,11 ]\}$. Vertices are ordered to keep dependent iterations of SpTRSV  and consecutive iterations SpMV next to each other.



\section{Experimental Results}



We compare the performance of sparse fusion to MKL~\cite{wang2014intel} and ParSy~\cite{cheshmi2018parsy}, two state-of-the-art tools that accelerate individual sparse kernels, which we call unfused implementations. Sparse fusion is also compared to  three fused implementations that we create. To our knowledge, sparse fusion is the first work that provides a fused implementation of sparse kernels where at least one kernel has loop-carried dependencies. For comparison, we also create three fused implementations of sparse kernels by applying  LBC, DAGP, and a wavefront technique to the joint DAG of the two input sparse kernels and create a schedule for execution using the created partitioning, the methods will be referred to as fused LBC, fused DAGP, and fused wavefront in order. 

\begin{table}
 \caption{The list of sparse matrices. 
 }

\begin{tabular}{  p{0.1cm}p{1.4cm}p{1.19cm} V{2}V{2} p{0.1cm}p{1.9cm}p{1.17cm} } 
\toprule
 \textbf{ID}&\centering Name &  Nonzeros & \textbf{ID} & \centering Name & Nonzeros \\
\rowcolor{tabcol}\textbf{1}& Flan\_1565 & 117.4$\times 10^6$ & \textbf{5} &Emilia\_923   &41$\times 10^6$ \\
 \textbf{2} & bone010 & 71.7$\times 10^6$ & \textbf{6} & StocF-1465 & 21$\times 10^6$ \\
\rowcolor{tabcol} \textbf{3} & Hook\_1498 & 60.9$\times 10^6$ & \textbf{7} & af\_0\_k101  &17.6$\times 10^6$ \\
\textbf{4}& af\_shell10 & 52.3$\times 10^6$ & \textbf{8} & ted\_B\_unscal  & 0.14$\times 10^6$  \\

\bottomrule
 \end{tabular}

 \label{tab:matrixlist}
\end{table}

\begin{table*}[!ht]
 \caption{The list of kernel combinations. CD: loops with carried dependencies,   SpIC0: Sparse Incomplete Cholesky with zero fill-in, SpILU0: Sparse Incomplete LU with zero fill-in, DSCAL: scaling rows and columns of a sparse matrix.  
 }
\begin{tabular}{ p{0.1cm}p{3.8cm}p{2.6cm}cc } 
\toprule
 ID & Kernel combination & Operations & Dependency DAGs &  Reuse Ratio \\
 \rowcolor{tabcol} 1 &\small SpTRSV CSR - SpTRSV CSR & \small $x=L^{-1}y, z=L^{-1}x$  &\small CD - CD &  $\frac{2n + 2size_L}{max(2n+size_L, size_L + 2n)} \ge 1$  \\
2 &\small SpMV CSR - SpTRSV CSR& \small $y = Ax,$ $z = L^{-1}y$ & \small Parallel - CD   &  $\frac{2n}{max(2n+size_L, size_A + 2n)} < 1$  \\
\rowcolor{tabcol}3 &\small DSCAL CSR - SpILU0 CSR & \small $LU \approx DAD^{T}$  &\small Parallel - CD &  $\frac{2size_A}{max(size_A, size_A + 2n)} \ge 1$  \\
4 &\small SpTRSV CSR - SpMV CSC& \small $y = L^{-1}x,$ $z = Ay$  &  \small CD - Parallel & $\frac{2n}{max(2n+size_L, size_A + 2n)} < 1$    \\
\rowcolor{tabcol}5 &\small SpIC0 CSC - SpTRSV CSC& \small $LL^{T} \approx A,$ $y = L^{-1}x$  &\small CD - CD &  $\frac{2size_L}{max(size_L, size_L + 2n)} \ge 1$  \\
6 &\small SpILU0 CSR - SpTRSV CSR& \small $ LU \approx A,$ $y = L^{-1}x$  &\small CD - CD &  $\frac{2size_A}{max(size_A, size_L + 2n)} \ge 1$  \\
\rowcolor{tabcol}7 &\small DSCAL CSC - SpIC0 CSC & \small $LL^{T} \approx DAD^{T}$  &\small Parallel - CD & $\frac{2size_L}{max(size_L, size_L + 2n)} \ge 1$  \\

\bottomrule
 \end{tabular}

 \label{tab:kernellist}
\end{table*}

\noindent\textbf{Setup.} 
The set of symmetric positive definite matrices listed in Table~\ref{tab:matrixlist} are used for experimental results.
The matrices are from~\cite{davis2011university} and with real values in double precision.
The test-bed architecture is a\hkaz{socket of XSEDE Comet ~\cite{xsede}} \zedka{multicore processor} with 12 cores of a Xeon E5-2680v3 processor with 30MB L3 cache. 
 All generated codes, implementations of different approaches, and library drivers are compiled with GCC v.7.2.0 compiler and with \rebt{the} \texttt{-O3} flag. 
Matrices are first reordered with METIS~\cite{karypis1998software} 
to improve parallelism.


We compare \DSL{} with two unfused implementations where each kernel is optimized separately: \emph{I. ParSy} applies LBC to DAGs that have edges. For parallel loops, the method runs all iterations in parallel. LBC is developed for L-factors~\cite{davis2006direct} or chordal DAGs. Thus, we make DAGs chordal before using LBC. \emph{II. MKL } uses Intel MKL~\cite{wang2014intel} routines with MKL 2019.3.199 and calls them separately for each kernel. 

Sparse fusion is also compared to three  fused approaches all of which take as input the \emph{joint DAG}; the joint DAG is created from combining the  DAGs of the input kernels using the inter-DAG dependency matrix $F$. We then implement three approaches to build the fused schedule from the joint DAG: \emph{I. Fused wavefront} traverses the joint DAG in topological order and builds a list of wavefronts that represent vertices of both DAGs that can run in parallel.
\emph{II. Fused LBC} applies the LBC algorithm to the joint DAG and creates a set of s-partitions each composed of independent w-partitions. 
Then the s-partitions are executed sequentially and w-partitions inside an s-partition are executed in parallel. LBC is taken from ParSy and its parameters are tuned for best performance. The joint DAG is first made chordal and then passed to LBC.  \emph{III. Fused DAGP} applies the DAGP partitioning algorithm to the joint DAG and then executes all independent partitions that are in the same wavefront in parallel. DAGP is used with METIS for its initial partitioning, with one run (\texttt{runs=1}) and the remaining parameters are set to default.

The list of sparse kernel combinations  investigated  are  in Table~\ref{tab:kernellist}. To demonstrate sparse fusion's capabilities, the sparse kernels are selected with different combinations of storage formats, i.e. CSR and compressed sparse column (CSC) storage,  different combinations of parallel loops and loops with carried dependencies, and a variety of memory access pattern behaviour. 
For example, combinations of SpTRSV, $Lx=b$ and SpMV are main bottlenecks in conjugate gradient methods ~\cite{zhuang2017iteration,benzi2000robust}, GMRES ~\cite{cheshmi2020nasoq}, Gauss-Seidel ~\cite{saad2003iterative}. Preconditioned Krylov methods ~\cite{grigori2015communication} and Newton solvers~\cite{soori2018reducing} frequently use kernel combinations 3, 5, 6, 7. The s-step Krylov solvers ~\cite{carson2015communication} and s-step optimization methods used in machine learning ~\cite{soori2018reducing} provide even more opportunities to interleave iterations. Thus, they use these kernel combinations significantly more than their classic formulations. 

\begin{table}[!ht]
 \caption{\rebt{The achieved GFLOP/s for the baseline code for the kernel combinations in Table~\ref{tab:kernellist} and for matrices in Table~\ref{tab:matrixlist}.  }
 }
\begin{tabular}{ ccccccccc } 

\toprule
    \multicolumn{8}{c}{Kernel Combination ID} \\  
 Matrix ID  & 1 & 2 & 3 & 4 & 5 & 6 & 7 \\
\rowcolor{tabcol}  1 & 1.52 & 1.54  & 0.45 & 1.55& 0.61& 0.43 & 0.61 \\
 2 & 1.5 & 1.54  & 0.45 & 1.54 & 0.61& 0.45 & 0.61  \\
\rowcolor{tabcol}  3 & 1.4 & 1.45  & 0.47 & 1.45 & 0.48 & 0.50 & 0.47  \\
 4 & 1.47 & 1.48  & 0.72 & 1.49 & 0.50 & 0.77 & 0.47  \\
\rowcolor{tabcol} 5 & 1.42 & 1.47  & 0.45 & 1.47 & 0.51 & 0.46 & 0.49  \\
 6 & 0.91 & 1.14  & 0.17 & 1.14 &  0.33 & 0.18 & 0.32  \\
\rowcolor{tabcol}  7 & 1.47 & 1.50  & 0.73 & 1.49 &  0.49 & 0.77 & 0.48  \\
 8 & 1.41 & 1.70  & 0.89 & 1.70& 0.44& 0.76 & 0.42  \\
\bottomrule
 \end{tabular}

 \label{tab:baseflops}
\end{table}

 \begin{figure*}
 \includegraphics[width=0.984\textwidth]{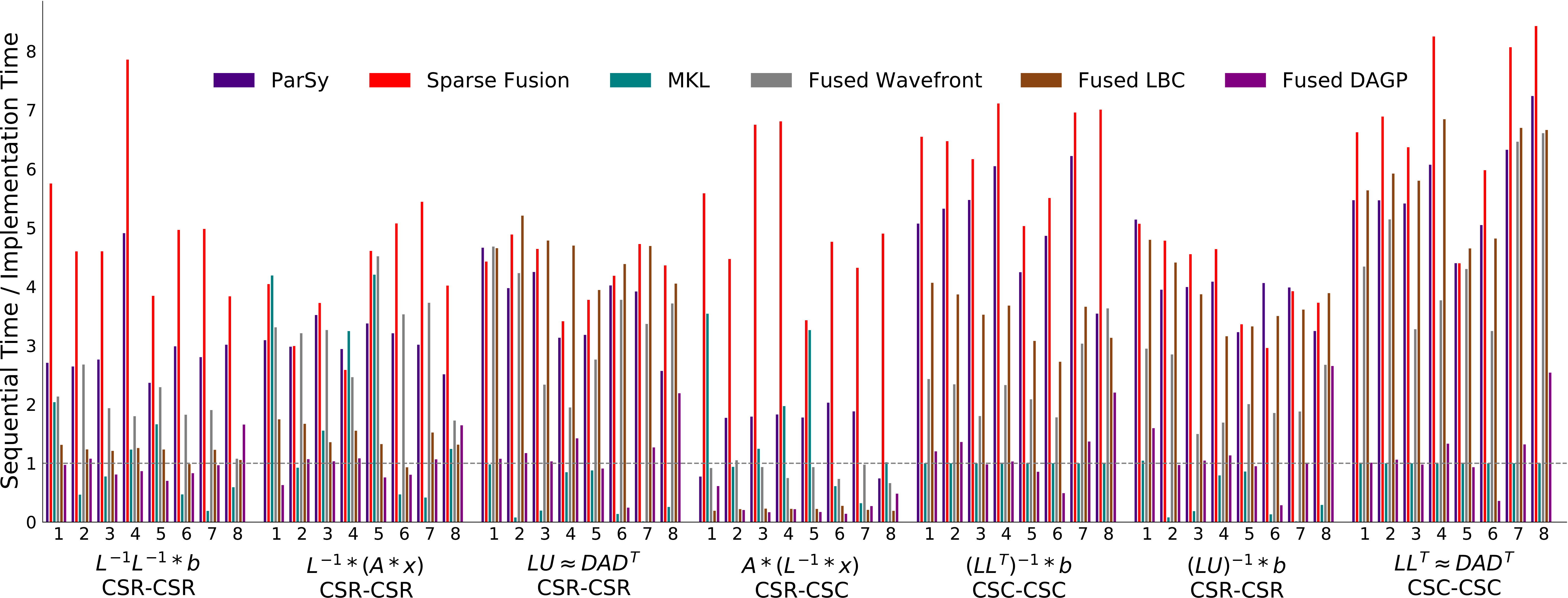}
 \caption{ Performance of different implementations shown with speedup from dividing baseline time by implementation time. 
 }
 \Description{fusion executor performance. }
 \label{fig:executor_all}
 \end{figure*}


\noindent\textbf{Sparse Fusion's Performance.} 
Figure~\ref{fig:executor_all} shows the performance of the fused code from  \DSL{}, the unfused implementation from ParSy and MKL, and the  fused wavefront, fused LBC, and fused DAGP implementations. All execution times are normalized over a \textit{baseline}.  The baseline  is obtained by running each kernel individually with a sequential implementation. \rebt{The floating point operations per second (FLOP/s) for each implementation can be obtained by multiplying the baseline FLOP/s from Table~\ref{tab:baseflops} with the speedups in Figure~\ref{fig:executor_all}. }
The \DSL's fused code is on average 1.6$\times$ faster than ParSy's executor code and 5.1$\times$ faster than MKL across all kernel combinations. Even though sparse fusion is on average 11.5$\times$ faster than MKL for ILU0-TRSV, since ILU0 only has a sequential implementation in MKL, the speedup of this kernel combination is excluded from the average speedups. The fused code from
\DSL{} is on average 2.5$\times$, 5.1$\times$, and 7.2$\times$ faster than in order fused wavefront, fused LBC, and fused DAGP. Obtained speedups of \DSL{} over ParSy (the fastest unfused implementation) 
for SpILU0-SpTRSV and SpIC0-SpTRSV is lower than other kernel combinations. Because SpIC0 and SpILU0  have a high execution time, when combined with others sparse kernels with a noticeably lower execution time, the realized speedup from fusion will not be significant.



\begin{figure}
 \includegraphics[width=0.45\textwidth]{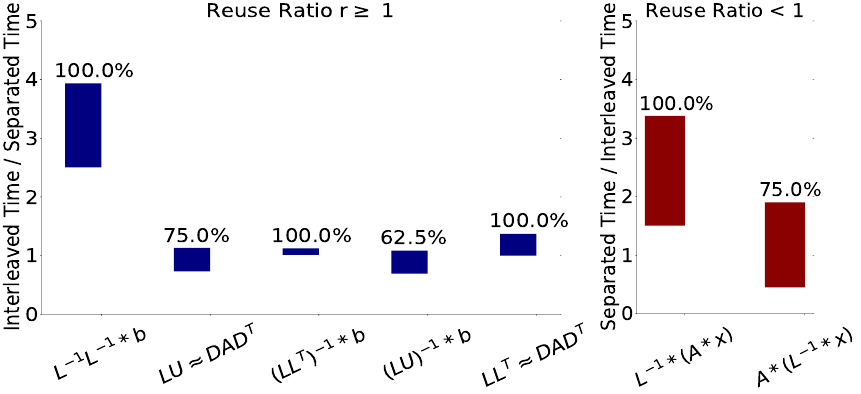}
 \caption{ The range of speedup for all matrices achieved as a result of using interleaved packing vs. separated packing. 
 The labels on bars  show how often the choice of packing strategy made by sparse fusion leads to performance improvement. }
 \Description{Profiling overall. }
 \label{fig:reuse}
 \end{figure}

\noindent\textbf{Locality in Sparse Fusion.} 
Figure~\ref{fig:reuse} shows the efficiency of the two packing strategies to improve locality. 
The effect of the packing strategy is shown for kernel combinations with a reuse ratio smaller and larger than one as shown in Table~\ref{tab:kernellist}.
Kernel combinations 1, 3, 5, 6, and 7 share the sparse matrix $L$ and thus have a reuse ratio larger than one while combination 2 and 4 only share vector $y$ leading to a reuse ratio lower than one.
Figure~\ref{fig:reuse} shows the range of speedup over all matrices for the selected packing strategy versus the other other packing method for each combination. 
As shown, the selected packing strategy in sparse fusion  improves the performance in 88\% of kernel combinations and matrices and provides 1-3.9$\times$  improvement in both categories.

\begin{figure}
 \includegraphics[width=0.45\textwidth]{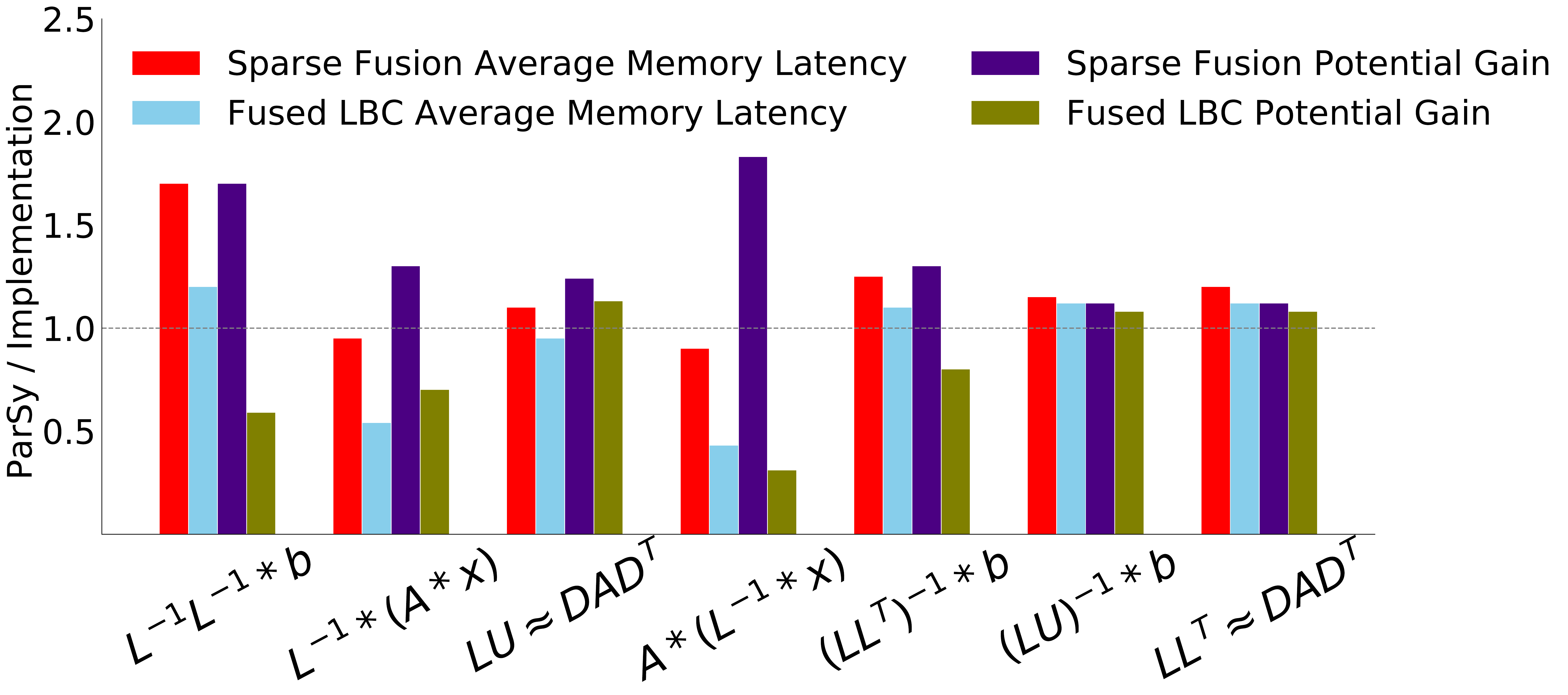}
 \caption{ Average memory access time and the OpenMP potential gain 
 for matrix \textit{bone010}. The legends show the implementation, values are normalized over  ParSy.
 }
 \Description{Profiling. }
 \label{fig:profiling}
 \end{figure}

Figure~\ref{fig:profiling} shows the average memory access latency~\cite{hennessy2017computer} of sparse fusion, the fastest unfused implementation (ParSy), and the fastest fused partitioning-based implementation (Fused LBC) for all kernel combinations normalized over the ParSy average memory access latency (shown for  matrix \textit{bone010} as example, other matrices  exhibit similar behavior). The average memory access latency is used as a proxy for locality and is computed using the number of accesses to L1, LLC, and TLB measured with PAPI performance counters~\cite{terpstra2010collecting}. 

 For kernels 1, 3, 5, 6, and 7 where the reuse ratio is larger than one, the memory access latency of ParSy is on average 1.3$\times$ larger than that of sparse fusion. Because of their high reuse ratio, these kernels benefit from optimizing locality between kernels made possible via interleaved packing. ParSy optimizes locality in each kernel individually. When applied to the joint DAG, LBC can potentially improve the temporal locality between kernels and thus there is only a small gap between the memory access latency of sparse fusion  and that of  fused LBC. For kernels 2 and 4 where the reuse ratio is smaller than one, the gap between the memory access latency of sparse fusion and fused LBC is larger than the gap between the memory access latency of sparse fusion and ParSy. Sparse fusion and ParSy both improve data locality within each kernel for these kernel combinations.

\noindent\textbf{Load Balance and Synchronization in Sparse Fusion.} %
%
Figure~\ref{fig:profiling} shows the OpenMP potential gain~\cite{potentialgainintel} of sparse fusion, ParSy, and Fused LBC for all kernel combinations normalized over ParSy’s potential gain (shown for  
matrix \textit{bone010} 
as example, but all other matrices in Table~\ref{tab:matrixlist} follow similar behavior.) 
The OpenMP potential gain is a metric in Vtune~\cite{zoneintel} that shows the total parallelism overhead, e.g. wait-time due to load imbalance and synchronization overhead, divided by the number of threads. This metric is used to measure the load imbalance and synchronization overhead in ParSy, fused LBC, and sparse fusion. 

Kernel combinations 2 and 4 have slack vertices that provide opportunities to balance workloads. For example, for matrices shown in Table~\ref{tab:matrixlist}, between 35-76\% vertices can be slacked thus the potential gain balance of ParSy is 1.6$\times$ larger than sparse fusion and 2.4$\times$ lower than fused LBC.   
ParSy can only improve load balance using the workloads of an individual kernel. 
As shown in Figure~\ref{fig:problem}, for the kernel combination 5, the joint DAG has a small number of parallel iterations in final wavefronts that makes the final s-partitions of the LBC fused implementation imbalanced (a similar trend exists for kernel combination 6). 
For these kernel combinations, the code from sparse fusion has on average 33\% fewer synchronization barriers compared to ParSy due to merging.
For kernel combinations 1, 2, 3, 4, and 7 the potential gain in sparse fusion is 1.3$\times$ less than that of ParSy. Merging in sparse fusion reduces the number of synchronizations in the fused code on average 50\% compared to that of ParSy.

\begin{figure}
 \includegraphics[width=0.475\textwidth]{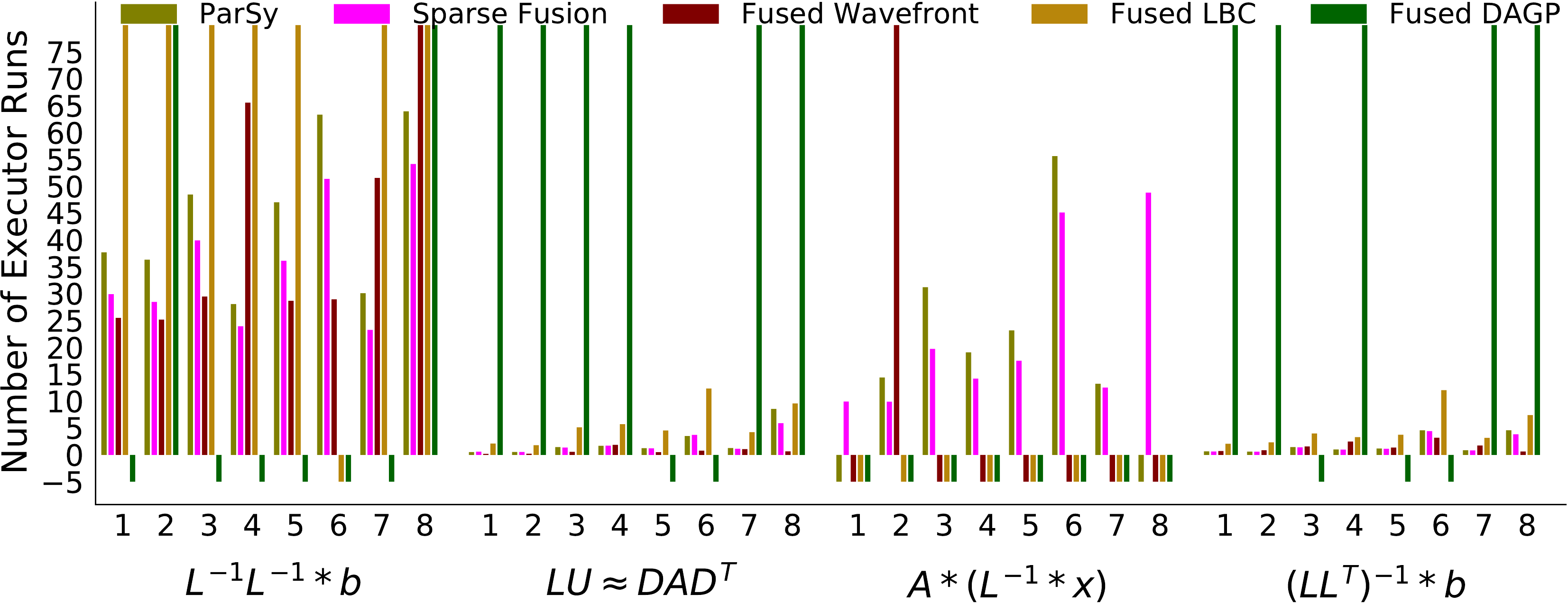}
 \caption{ The number of executor runs to amortize  inspector cost. 
 Values are clipped between -5 and 80. 
 (lower is better)}
 \Description{fusion inspector performance. }
 \label{fig:inspector_all}
 \end{figure}




\noindent\textbf{Inspector Time.}
Figure~\ref{fig:inspector_all} shows the number of times that the executor should run to amortize the cost of inspection for implementations that have an inspector. \kazed{For space only combinations 1, 3, 4, and 5 are shown, others follow the same trend.}  
The number of executor runs (NER) that amortize the cost of inspector for an implementation  is calculated using \\$ \frac{ \>\> Inspector \>\> Time}{Baseline\> Time -  \>\> Executor\>\> Time} $. The \textit{baseline} time is obtained by running each kernel individually with a sequential implementation, the inspector and executor times belong to the specific implementation.
The fused LBC implementation has a NER of 3.1-745. The high inspection time is because of the high  cost of converting the joint DAG into a chordal DAG, typically consuming 64\% of its inspection time.    
The NER of the fused DAGP implementation is either negative or higher than 80. 
The fused wavefront implementation sometimes has a negative NER because the executor time is slower than the baseline time.
As shown, \DSL{} and fused wavefront have the lowest NER amongst all implementations. 
Sparse fusion's low inspection time is due to pairing strategies that enable partitioning one DAG at a time.
Kernel combinations such as, SpIC0-TRSV and SpILU0-TRSV only need one iteration to amortize the inspection time and  SpTRSV-SpMV, SpTRSV-SptRSV, and SpMV-SpTRSV need between 11-50 iterations. 
Sparse kernel combinations  are routinely used in iterative solvers in  scientific applications. Even with preconditioning, these solvers typically converge to an accurate solution after ten of thousands of iterations \cite{benzi2000robust,kershaw1978incomplete,papadrakakis1993accuracy}, hence amortizing the overhead of inspection.
\section{Related work}
Parallel implementations of individual sparse matrix kernels exist in both highly-optimized libraries~\cite{henon2002pastix,li2005overview}
and inspector-executor approaches~\cite{cheshmi2017sympiler,mohammadi2019sparse,strout2018sparse}. Some libraries such as MKL~\cite{wang2014intel}, and code generators such as Taichi~\cite{10.1145/3355089.3356506} and TACO~\cite{kjolstad2017taco} provide optimizations for a range of sparse matrix kernels, while others provide optimizations for a specific sparse kernel. For example, the sparse triangular solve has been optimized in~\cite{li2013gpu,naumov2011parallel,wang2018swsptrsv,vuduc2002automatic,totoni2014structure,yilmaz2020adaptive,Park2014,Picciau2016,LevelSet90}, optimizations of sparse matrix-vector multiply are available in~\cite{williams2009optimization,kamin2014optimization,merrill2016merge,li2013smat,ashari2014fast,liu2018towards}, and LU and Cholesky factorization have been optimized in SuperLU~\cite{li2005overview} and Pastix~\cite{henon2002pastix}.  

Inspector-executor frameworks commonly use wavefront parallelism~\cite{venkat2016automating,rauchwerger1995run,zhuang2009exploiting,strout2002combining,naumov2011parallel,govindarajan2013runtime} to parallelize sparse matrix computations with loop-carried dependencies. 
Recently, task coarsening approaches such as LBC~\cite{cheshmi2018parsy} and DAGP~\cite{herrmann2019multilevel} 
coarsen wavefronts and thus generate code that is optimized for parallelism, load balance, and locality. While available approaches can provide efficient optimizations for sparse kernels with or without loop-carried dependencies, they can only optimize sparse kernels individually. 


A number of libraries and inspector-executor frameworks provide parallel implementations of fused sparse kernels with no loop-carried dependencies such as,
two or more SpMV kernels\rebt{~\cite{hoemmen2010communication,6514719,6375534,aliaga2015systematic,rupp2016viennacl}} or SpMV and dot products~\cite{10.1145/3079079.3079091,dehnavi2011enhancing,aliaga2015systematic,ghysels2014hiding,agullo2009numerical,rupp2016viennacl}. The formulation of $s$-step Krylov solvers~\cite{carson2015communication} has enabled iterations of iterative solvers to be interleaved and hence multiple SpMV kernels are optimized simultaneously via replicating computations to minimize communication costs~\cite{hoemmen2010communication,6514719,6375534,soori2018reducing}. 
Sparse tiling~\cite{strout2003compile,krieger2013loop,strout2014generalizing,strout2002combining,strout2004sparse} is an inspector executor approach that uses manually written inspectors~\cite{strout2003compile,strout2004sparse} to group iteration of different loops of a specific kernel such as Gauss-Seidel~\cite{strout2004sparse} and Moldyn~\cite{strout2003compile} and is generalized for parallel loops without loop-carried dependencies~\cite{strout2014generalizing,krieger2013loop}.  Sparse fusion optimizes combinations of sparse  kernels where at least one of the kernels has loop-carried dependencies. 

\section{Conclusion}
We present sparse fusion  and demonstrate how it improves parallelism, load balance, and data locality in sparse matrix combinations compared to when sparse kernels are optimized separately. Sparse fusion inspects the DAGs of the input sparse kernels and uses the MSP algorithm to balance the workload between wavefronts and determine whether to optimize data locality for within or between the kernels. Sparse fusion's generated code outperforms state-of-the-art implementations for sparse matrix optimizations. 
In future work, we plan to investigate strategies that select the most profitable loops to  be fused to support the fusion of more than two loops.

\begin{acks}                            
This work was supported in part by NSERC Discovery Grants (RGPIN-06516, DGECR00303), the Canada Research Chairs program, and U.S. NSF awards NSF CCF-1814888, NSF CCF-1657175; used the Extreme Science and Engineering Discovery Environment (XSEDE) [Towns et al. 2014] which is supported by NSF grant number ACI-1548562; and was enabled in part by Compute Canada and Scinet \footnote{www.computecanada.ca}.
\end{acks}

\balance
\bibliography{syppRef}



\end{document}